\documentstyle[seceq,preprint]{jpsj}

\def\runtitle{
Renormalized Perturbation Approach  for Itinerant-Localized Model
}
\def\runauthor{Yukihiro {\sc Okuno}, Osamu {\sc Narikiyo} and Kazumasa {\sc Miyake}}

\title
{
Renormalized Perturbation Approach  for Examination of 
Itinerant-Localized Duality Model for Strongly  Correlated Electron Systems
}

\author
{
Yukihiro {\sc Okuno}\footnote{e-mail: okuno@aquarius.mp.es.osaka-u.ac.jp}, 
Osamu {\sc Narikiyo}${^1}$ and Kazumasa {\sc Miyake}${^1}$
}

\inst
{
Department of Physics, Faculty of Science,
Kyoto University, Kyoto 606 \\
${^1}$Department of Material Physics, Faculty of Engineering Science,
Osaka University, Toyonaka 560
}

\recdate
{March 11, 1997
}

\abst
{We present a microscopic examination for the 
 itinerant-localized duality model which has 
been proposed to understand anomalous properties of 
  strongly correlated systems like the heavy fermions by Kuramoto
 and Miyake, and also useful to describe the anomalous properties of 
 the high-$T_{c}$ cupurates.
 The action of the duality model consists of
 two components of dynamical degrees of freedom.
 One of them is coherent itinerant degree of fermion and the other is
 incoherent localized one describing mainly the
 spin which has not 
 been  explicitly  considered so far.
  We show that the thermodynamic potential of the strongly interacting 
 Hubbard model can be rearranged in the form of duality model
 on the basis of renormalized perturbation expansion of the 
 Luttinger-Ward functional if the one-particle spectral weight 
 exhibits triple peak structure.
 We also examine the incoherent degrees of freedom described 
 as a ``localized spin''  and show on the basis of the pertubation expansion
 that there exists commensurate 
 superexchange-type interaction among the ``localized spins''.
}

\kword
{
itinerant-localized duality model, strongly correlated systems, superexchange
interaction
}

\begin{document}
\sloppy
\maketitle

\section{Introduction}
 The Fermi liquid theory offers us  a very powerful tool for describing the low
 energy properties of strongly correlated fermions.
 While it was first devised by Landau~\cite{rf:Landau} to analyse the physics
 of liquid $^{3}$He, its general and profound value has been recognized through
 the application to the many areas of correlated fermion systems such as Kondo
 problem for instance.~\cite{rf:Nozieres,rf:Yamada} 
 But it is also true that there are many phenomena which at a first glance
 cannot be explained by applying 
 the Fermi liquid theory in its simple form. For example, heavy
 fermions and the high T${}_{c}$ cuprates 
 offer such phenomena.
 In the former they are the metamagnetic behavior and the extremely weak
 antiferromagnetism in the metallic phase. In the latter case 
 they are ``spin-gap'' behavior and ``spin-charge separation aspect'',
 and the controversy about the applicability and limitation of 
 the Fermi liquid theory have been arisen.
 Itinerant-localized duality model, which has been 
 proposed as a quantum 
 phenomenology by Kuramoto and Miyake,~\cite{rf:Kuramoto1} 
 looks for the origin of such phenomena with the intermediate energy
 scale in an explicit treatment of the incoherent
 contribution which is renormalized in quasiparticle in the ordinary 
 Fermi liquid theory. It 
 has been  applied not only to the heavy 
 fermions systems,~\cite{rf:Kuramoto2,rf:Miyake1} but also the 
 understanding the anomalous properties of high-{\it T}${}_{c}$
 cuprates.~\cite{rf:Miyake2,rf:Narikiyo1,rf:Narikiyo2} \\
 \indent This model is based on the physical picture that, 
 through the renormalization group evolution to the fixed point
 of such strong correlated electron systems, there is an
 intermediate stage where both the itinerant degree of freedom 
 and ``localized-spin'' degree of freedom become apparent and interact
 with each other.
 The physical ground of such a picture is attributed to the fact  
 that the one-particle
 spectral weight has triple peak structure in strongly 
 correlated fermion systems in general 
 and large amount of the spectral weight is transferred to
 the high energy incoherent part while that of the coherent one decreases
 with the absolute value at the Fermi level unchanged.
 The triple peak structure consist of two broad peaks corresponding to 
 the upper and lower Hubbard bands~\cite{rf:Hubbard} and the narrow
 quasiparticle peak of Fermi liquid accessible by 
 Gutzwiller's type description.~\cite{rf:Gutzwiller} 
 It has  been explicitly recognized
 by Kawabata~\cite{rf:Kawabata} two decade ago and demonstrated by
 recent numerical works 
 with $d$=$\infty$ technique, $d$ being the 
space dimension.~\cite{rf:Georges,rf:Jarrell,rf:Zhang,rf:Pruschke,rf:Sakai}. \\
 \indent The incoherent part has influence on the physical quantities 
 of the systems even in the low energy region
 where the Fermi liquid description is valid. 
 In particular, the spin fluctuation spectrum
 is affected considerably  by the incoherent part,
 or ``localized spins'', through
 a diffusive motion of clusters of
 ``localized spins'' with short range antifferomagnetic order. 
 This is a salient nature of the duality
 model distinguished from a simple Fermi liquid description
 where the incoherent part is considered to be a simple
 featureless back ground.
 A possible importance of the incoherent parts was recognized
 by Ruckenstein and Varma~\cite{rf:Ruckenstein}
 in the argument for a microscopic basis of the marginal 
 Fermi liquid.
 We here retain the incoherent part explicitly
  as a ``localized spin'' degree
 of freedom which  is the irrelevant operator that should be renormalized
 into the vertex part of
 the itinerant fermion at the fixed point. \\
 \indent In this paper, we present a microscopic justification of the 
 duality model action for the Hubbard model with strong correlation
 on the basis of the renormalized perturbation expansion of the
 Luttinger-Ward skeleton representation for the thermodynamic potential.
 In this process, we make explicit use of the triple peak structure
 of the one-particle spectral weight.
 The organization of this paper is as follows. 
 In $\S2$ we develop a microscopic
 justification of the duality model action 
 and in $\S3$
 we investigate the role of the high energy process on the
 fluctuations of the ``localized spins'' which have commensurate correlation
 through a superexchange-type interaction. Conclusion and discussion 
 are given in $\S4$.
\section{Justification of duality model 
 - separation of itinerant and localized degree of freedom - }
The effective action of the duality model 
for low energy physics of strongly correlated 
electron systems is written as ~\cite{rf:Kuramoto1}
\begin{eqnarray}
{\it A} &=& A_{{\rm f}} + A_{{\rm s}} +A_{{\rm int}},
\end{eqnarray}
where
\begin{subeqnarray}
 A_{{\rm f}} &=& 
 -\sum_{i,j.\sigma}\sum_{n}f^{\dagger}_{i \sigma}(-{\rm i}\epsilon_{n})
 (\bar{G}^{-1}_{ij,\sigma}
({\rm i}\epsilon_{n}))f_{j \sigma}({\rm i}\epsilon_{n}), \\
 A_{{\rm s}} &=& 
 \frac{1}{2}\sum_{i,j,m}{\bf S}{}_{i}(-{\rm i} \nu_{m})
(\chi_{0 ij}^{-1}({\rm i}\nu_{m})){\bf S}{}_{j}
({\rm i} \nu_{m}), \\
 A_{{\rm int}} &=& 
-\lambda_{0}\sum_{i \alpha \beta}\sum_{mn}f^{\dagger}_{i \alpha}
(-{\rm i}\epsilon_{n}-{\rm i}\nu_{m})
f_{i \beta}({\rm i}\epsilon_{n}){\bf \sigma}{}_{\alpha \beta}
\cdot{\bf S}{}_{i}({\rm i}\nu_{m}),
\end{subeqnarray}
 where $A_{{\rm f}}$ and $A_{{\rm s}}$ represent the parts of fermions
 and localized spin consisting of incoherent part of fermions respectively, 
 and $A_{{\rm int}}$ is the interaction term of 
 the fermion with the localized spin. 
 The spin degree of freedom, ${\bf S}_{i}(\nu_{m})$,
 mainly comes from the particle-hole exicitations with 
 high energy part of fermions, corresponding to the Hubbard bands.
 However, energies ${\rm i}\nu_{m}$ of ${\bf S}_{i}$ are restricted 
 essentially within the low-energy region comparable to the 
 renormalized bandwidth of itinerant fermions.
 This is assumed by the structure of the dynamical susceptibility 
 $\chi_{0ij}({\rm i}\nu_{m})$ as the energies 
${\rm i}\epsilon_{n}$ of fermions 
 are restricted by the structure of $\bar{G}_{ij}({\rm i}\epsilon_{n})$.
 It is also noted that ${\bf S}_{i}$ is a boson field describing the 
 quantum fluctuation of the ``localized spins'' which has been almost quenched
 by high energy process such as the Kondo effect in Ce-based heavy fermions.\\
 \indent What we want to show below is that the  action
(2.1) is actually appropriate form for strongly correlated 
 electron systems in the low energy region.
 If the renormalization-flow line of the original Hamiltonian of
 strongly correlated electron systems 
 passes through the point where the duality model 
 action can apply in the vicinity of
 the fixed point, then this duality action is a correct 
 effective action of the original one.
 So we try to show that in the vicinity of the fixed point
 the renormalization drives the original Hamiltonian,
 which has strong  on-site Coulomb interaction, toward the form of the duality
 model action. \\
 \indent A strategy we take here is to show that the thermodynamic potential
 in the duality model and that of the
 original Hamiltonian actually coincide with each other term by term under
 a certain condition satisfied by strongly
 correlated electron systems. The duality model is the effective model 
 in the region where the renormalization has been carried out 
 appreciably but not completely.
 So we take the Luttinger--Ward functional form which is the functional
 of the fully renormalized Green function and the self energy 
 of the original Hamiltonian
 and expand it about the point corresponding to the duality model.
 The thermodynamic potential ${\it Y}$  of the original Hamiltonian of
 the Luttinger-Ward~\cite{rf:Luttinger} form is
 \begin{eqnarray}
 {\it Y} &=& T\sum_{\epsilon_{n}}\sum_{p,\sigma}e^{{\rm i}\epsilon_{n}\delta}
 [\ln G_{\sigma}(p,{\rm i}\epsilon_{n})
 -\Sigma_{\sigma}(p,{\rm i}\epsilon_{n})G_{\sigma}(p,{\rm i}\epsilon_{n})]+
 \Phi [G_{\sigma}(p,{\rm i}\epsilon_{n})],
 \end{eqnarray}
 where $G_{\sigma}(p,{\rm i}\epsilon_{n})$ and 
$\Sigma_{\sigma}(p,{\rm i}\epsilon_{n})$ 
 is fully renormalized Green function
 and self energy respectively and $\Phi [G_{\sigma}(p,{\rm i}\epsilon_{n})]$ 
 is the Luttinger-Ward functinal defined by
\begin{eqnarray}
 \Phi [G_{\sigma}] &=& T\sum_{\epsilon_{n}}\sum_{p,\sigma}\sum_{n} 
 \frac{1}{2n}\Sigma^{n}{}_{\sigma}(p,{\rm i}\epsilon_{n})
 G_{\sigma}(p,{\rm i}\epsilon_{n}),
\end{eqnarray}
 where $\Sigma^{n}{}_{\sigma}(p,{\rm i}\epsilon_{n})$ is the 
 {\it n}-th order contribution to the self energy from
 all the possible skeleton diagrams with
 the fully renormalized Green functions $G_{\sigma}(p,{\rm i}\epsilon_{n})$. 
 To avoid an overcounting
 of the same diagram composed of 2$n$ Green function, 
 the factor $\frac{1}{2n}$ appears in (2.4).\\
 \indent We show the expansion of the Luttinger-Ward functional 
 $\Phi [G_{\sigma}]$ up to  the 4th order term of the Hubbard model 
 in Fig.[1],
 the solid line is the renormalized Green function and the dotted 
 line is the bare Coulomb interaction $U$.
 In Fig.[2], on the other hand, we also show the expansion of 
 thermodynamics potential $\Omega$
 of the duality model up to 4th order of coupling constant $\lambda_{0}$. 
 In Fig.[2],
 the solid line represents the Green function $\bar{G}_{\sigma}$ 
 of the fermion of the duality model 
 and the wavy line the propageter $\chi_{0}$ of the ``localized spin'', 
 and $\Omega_{0}$ is the noninteracting part of the thermodynamical potential
 which is written as
\begin{eqnarray}
 \Omega_{0} &=& T\sum_{\epsilon_{n}}\sum_{p,\sigma}
 e^{{\rm i}\delta\epsilon_{n}}\ln\bar{G}{}_{\sigma}(p,{\rm i}\epsilon_{n}).
\end{eqnarray}
 \indent If the expansion (2.3) of $Y$, the thermodynamic potential of
 the original Hamiltonian, is reduced
 to the expansion of $\Omega$, that of the duality model, 
 under a certain reasonable 
 condition, we get the
 correctness of the duality model action (2.1) $\sim$ (2.4). 
 To show the equivalence of  the expansion $\Omega$
 to ${\it Y}$, the diagrams of ${\it Y}$ need to be reduced to those of 
 $\Omega$ with correct coefficients of corresponding diagrams.
 The expansion of the thermodynamic potential ${\it Y}$ contains the fully
 renormalized Green function $G_{\sigma}$ which
 describes the physics around the fixed point. 
 On the other hand, the expansion of $\Omega$ contains the Green function
 of duality model $\bar{G}_{\sigma}$ which is only partially renormalized. 
 The fully renormalized
 Green function $G$ is expanded as follow.
\begin{eqnarray}
 G_{\sigma}(p,\epsilon) &=& \tilde{G}{}_{\sigma}(p,\epsilon)+\delta G_{\sigma}(p,\epsilon),
\end{eqnarray}
 where $\tilde{G}{}_{\sigma}(p,\epsilon)$ is the partially renormalized 
 Green function appearing in the 
 duality model action (2.2a) and $\delta G_{\sigma}(p,\epsilon)$ is the 
 correction due to the
 renormalization by the interaction (2.2c). Here we set up 
 following physical assumptions.\\
\indent First we assume that in the energy region, where the 
 duality model can apply, 
 the coherent fermions interact only through the spin fluctuations because
 the charge fluctuations with high energy scale are suppressed 
 in strongly correlated electron systems.
 The diagrams of $\delta G_{\sigma}(p,{\rm i}\epsilon_{n})$ are 
 given as shown in Fig.[3] 
 where the dotted line represents the partially renormalized
 Green function $\tilde{G}{}_{\sigma}(p,{\rm i}\epsilon_{n})$ and the 
 wavy line the spin fluctuations propagator
 $\chi_{0}$ and the dot the fermion-spin
 vertex part $\lambda_{0}$.\\ 
 \indent Another and the most crucial assumption is that
 the one-particle spectral weight given by $\tilde{G}_{\sigma}$ 
 already has the triple peak structure in the energy region in question.
 Namely, the quasi-particle
 peak is narrowed and  considerable part of the spectral 
 weight are transferred 
 to the high energy region of upper and
 lower Hubbard band broad peak. 
 The triple peak structure is a characteristic feature 
 of the strongly correlated
 electron systems. Due to this structure
 we can explicitly distinguish the high energy
 processes of fermion excitations  which belong to
 the incoherent parts of the upper and lower Hubbard bands 
 and the low energy processes
 of fermions which belong to
 the coherent part of quasiparticles. 
 The information that the electron system 
 is in strongly correlated region
 is condensed to the explicit separation of high and low energy part of fermion
 excitations.
 We can assume this {\it a posteriori}. \\
 \indent Hereafter we consider the diagrammatic structures of the Green
 function $\tilde{G}{}_{\sigma}(p,\epsilon)$ 
 and $\delta G_{\sigma}(p,\epsilon)$ in detail.
 The partially renormalized Green function $\tilde{G}{}_{\sigma}(p,\epsilon)$ 
 is written as follow
\begin{eqnarray}
 \tilde{G}^{-1}{}_{\sigma}(p,{\rm i}\epsilon_{n}) &=& 
 G^{-1}{}_{0 \sigma}(p,{\rm i}\epsilon_{n})
 -\tilde{\Sigma}{}_{\sigma}(p,{\rm i}\epsilon_{n}),
\end{eqnarray}
 where $G_{0 \sigma}(p,{\rm i}\epsilon_{n})$ is the bare Green function 
 and $\tilde{\Sigma}{}_{\sigma}(p,{\rm i}\epsilon_{n})$
 is a partially renormalized selfenergy 
 which is composed only of high energy process.
 Here the
 one particle-spectrum, expressed by
$ \rho (x) = -\frac{1}{\pi}{\rm Im}\tilde{G}{}_{\sigma}(p,x)$,
 has triple peak structure already and the low energy
 coherent parts are narrowed while the high energy
 incoherent parts are broadened.
 We can separate the high energy incoherent process and 
 the low energy coherent process in the spectral representation as
\begin{subeqnarray}
 \tilde{G}{}_{\sigma}(p,{\rm i}\epsilon_{n}) &=& 
 \int {\rm d}x \frac{\rho (x)}{x-{\rm i}\epsilon_{n}} \\
 &=& \int {\rm d}x \frac{\rho_{{\rm H}}(x)}{x-{\rm i}\epsilon_{n}} 
 + \int {\rm d}x \frac{\rho_{{\rm L}}(x)}{x- {\rm i}\epsilon_{n}} \\
 & \equiv & \tilde{G}^{{\rm H}}{}_{\sigma}(p,{\rm i}\epsilon_{n})
 +\tilde{G}^{{\rm L}}{}_{\sigma}(p,{\rm i}\epsilon_{n}).
\end{subeqnarray}
 The $\rho_{{\rm H}}(x)$ is the incoherent part of the one particle 
 spectral weight and
 the $\rho_{{\rm L}}(x)$ is the coherent one.
 The Green function $\tilde{G}^{{\rm H}}{}_{\sigma}(p,{\rm i}\epsilon_{n})$ 
 and the $\tilde{G}^{{\rm L}}{}_{\sigma}(p,{\rm i}\epsilon_{n})$  
 are defined as
\begin{subeqnarray}
 \tilde{G}^{{\rm H}}{}_{\sigma}(p,{\rm i}\epsilon_{n}) &=& 
 \int {\rm d}x \frac{\rho_{{\rm H}}(x)}{x-{\rm i}\epsilon_{n}}, \\
 \tilde{G}^{{\rm L}}{}_{\sigma}(p,{\rm i}\epsilon_{n}) &=& 
 \int {\rm d}x \frac{\rho_{{\rm L}}(x)}{x-{\rm i}\epsilon_{n}} \\
 &=& \frac{ \tilde{z}}{{\rm i}\epsilon_{n}-\tilde{\xi_{p}}},
\end{subeqnarray}
 where $\tilde{z}$ is the renormalization factor of 
 itinerant part of duality model, and is given
 as $\tilde{z}=(1-{\partial\tilde{\Sigma}(x)}/{\partial x})^{-1}$ 
 and the $\tilde{\xi_{p}}$ is a partially renormalized dispersion. 
 We note that the $\tilde{z}$ becomes already small 
 comparable to the order of $z$, the renormalization 
 factor at the fixed point.\\
\indent Now let us discuss a character of the spin fluctuations which
 are included in $\delta{G}{}_{\sigma}(p,{\rm i}\epsilon)$. 
 Due to the above assumptions, the spin fluctuations,
 composed of the incoherent part $\tilde{G}^{{\rm H}}{}_{\sigma}
(p,{\rm i}\epsilon)$,
 dominate in the energy region where the duality model can apply.
 The incoherent parts of fermion are included in the irrelevant
 operator in general.
 If our standing point is the fixed
 point of the renormalization, the irrelevant operators
 of the high energy process are renormalized into
 the vertex parts.
 However once we leave the fixed point, the irrelevant 
 operators will appear no matter how
 small they are, and in strongly correlated electron systems the
 spin fluctuation processes will be the largest among 
 such irrelevant operators.
 Considering the fact that the particle-hole 
 excitation with high energy fermions
 can carry the low energy, the spin fluctuation process from the incoherent
 part play an appreciable role for the interaction between fermions due to
 the broad upper and the lower Hubbard bands. From the Ward-identity 
 argument~\cite{rf:Kuramoto1} about the strength of spin-fermion 
 coupling for the impurity Anderson model, we can assume that fermions 
 strongly interact with the ``localized-spin''.
 So the diagrams of $\delta G$ in Fig.[3] can be regarded as we
 explicitly take out the irrelevant spin-fluctuation processes from the 
 selfenergy and neglect other fermion processes 
 owing to their relative smallness
 compared to the spin fluctuations.\\
\indent By substituting the Green function $G$, (2.6),
the thermodynamic potential $Y$, (2.3), can be rearranged as follows:
\begin{subeqnarray}
 Y &=& \ln G - \Sigma G + \Phi[G] \noindent \\
   &=& \ln\tilde G \\
   &+& \ln (1+\frac{\delta G}{\tilde{G}}) \\
   &-& (\tilde{\Sigma}+\delta\Sigma)(\tilde{G}+\delta G) \\
   &+& \Phi [\tilde{G}{}_{{\rm H}}] + 
\frac{\delta \Phi}{\delta G}|{}_{G=\tilde{G}{}_{{\rm H}}}
\delta^{\prime}G + \frac{\delta {}^{2} \Phi}{\delta G_{1} \delta G_{2}}|{}_{G=\tilde{G}{}_{{\rm H}}}
\delta^{\prime}G_{1}\delta^{\prime}G_{2} +\cdots ,
\end{subeqnarray}
 where $\delta^{\prime}G\equiv\tilde{G}{}_{{\rm L}}+\delta{G},$ and
 the summation with respect to frequency, momentum and spin
 are abbreviated.
 The contribution $\delta\Sigma$ represents the excess renormalization 
 on the basis of 
 the duality model in which only 
 $\tilde{\Sigma}$ has been taken into account. \\
\indent The term (2.10b) can be decomposed as
\begin{eqnarray}
 T\sum_{\epsilon_{n}}\ln\tilde{G}({\rm i}\epsilon_{n}) &=& 
 \int_{x\in {\rm H}} \frac{{\rm d}x}{2\pi}{\rm th}\frac{x}{2T}
 \tan^{-1}\frac{{\rm Im}\tilde{G}(x)}{{\rm Re}\tilde{G}(x)}
 + \int_{x\in {\rm L}} \frac{{\rm d}x}{2\pi}{\rm th}\frac{x}{2T}
 \tan^{-1}\frac{{\rm Im}\tilde{G}(x)}{{\rm Re}\tilde{G}(x)},
\end{eqnarray}
 where the symbol $x\in {\rm L}$ and $x\in {\rm H}$ 
 mean that the region of integration is restricted to the low and high
 energy region, respectively. The second term of (2.11)  
 includes only the low energy
 process and is equivalent to  
 $\ln\tilde{G}{}_{{\rm L}}$ which corresponds to 
 the thermodynamic potential $\Omega$, (2.5),
 the non-interacting part of
 the duality model, if $\bar{G}$ is identified with 
 $\tilde{G}{}_{{\rm L}}$. \\
\indent The term (2.10-c) is further expanded as
\begin{eqnarray}
\ln(1+\frac{\delta G}{\tilde{G}}) &\simeq& 
 \sum_{n=1}^{\infty}\frac{(-1)^{n-1}}{n}
( \frac{\delta G}{\tilde{G_{\rm L}}} ){}^{n}.
\end{eqnarray}
 where we have replaced $\tilde{G}$ by $\tilde{G}_{{\rm L}}$
in the right hand side because of the same reason as 
will be discussed below (2.14).
The first and the second term of (2.12) are shown in Fig.[4]. \\
\indent We next rearrange the term (2.10-d) as
\begin{eqnarray}
(\tilde{\Sigma}+\delta\Sigma)(\tilde{G}+\delta G) &=&
\tilde{\Sigma}\tilde{G}_{{\rm H}} + \tilde{\Sigma}(\delta^{\prime}G)
+ \delta\Sigma \tilde{G}{}_{{\rm H}} + \delta\Sigma \tilde{G}{}_{{\rm L}}
+ \delta\Sigma\delta G.
\end{eqnarray}
 The third term of (2.13) is the convolution of $\delta\Sigma$ 
 with the low energy components
  and with the $\tilde{G_{\rm H}}$ high energy ones, and is estimated as
\begin{eqnarray}
 T\sum_{{\epsilon_{n}}}\delta\Sigma ({\rm i}\epsilon_{n})
\tilde{G}_{{\rm H}}({\rm i}\epsilon_{n}) &=&
 \int \frac{{\rm d}x}{\pi} \int \frac{{\rm d}x^{\prime}}{\pi}
 \frac{{\rm Im}\delta\Sigma (x)
 {\rm Im}\tilde{G}{}_{{\rm H}}(x^{\prime})}{x-x^{\prime}}[f(x)-f(x^{\prime})],
\end{eqnarray}
 where $f(x)$ is the Fermi distribution function. 
 The integration value $x$ belongs to the coherent
 low energy region and  $x^{\prime}$ to the incoherent high energy region.
 Owing to the triple peak structure of one particle spectrum, 
 the denominator $x-x^{\prime}$ is roughly estimated about half of the on-site
 Coulomb energy $U$ in the main contribution of the integral, so this term is
 negligible. The first term of (2.13) gives a constant in the sense of the 
 duality model because it comes from the high energy processes.\\
\indent Finally, we consider the term (2.10-e). Here the equality
\begin{equation}
\frac{\delta \Phi}{\delta G}|{}_{G=\tilde{G}{}_{{\rm H}}} = \tilde{\Sigma}
\end{equation}
 holds because of the property of the functional $\Phi[G]$, 
 so that the second  term of (2.10e) cancels with the
 second term of (2.13).
 The remaining terms of (2.13) are shown in Fig.[5].
 The remaining terms of (2.10e) are obtained by replacing 
 $\tilde{G}_{{\rm H}}$'s in the
skeleton diagrams of $\Phi[G_{H}]$ with 
 $\tilde{G}_{{\rm L}}$ or 
$\delta G$ by the number of the order of functional derivative of $\Phi[G]$
 in all possible ways.
 On replacing $\tilde{G}_{{\rm H}}$ of the 
 $\Phi[G_{{\rm H}}]$ with $\tilde{G}_{{\rm L}}$ or $\delta G$,
 we select the diagrams which can contain   
 the spin fluctuations made of $\tilde{G}_{{\rm H}}$, representing 
 high energy process, and discard the
 diagrams from which we cannot extract the spin 
 fluctuation process as irrelevant ones
 in the energy region where the duality model can be applied.
 For example, of the diagrams shown in  Fig.[1] of 
 $\Phi [G]$ , c and f which contain only the singlet pair
fluctuations are regarded as irrelevant ones and will be discarded.
We assign the high energy incoherent part $\tilde{G}_{{\rm H}}$
for the spin fluctuation part of the diagrams of $\Phi [G]$
and the spin-fermion vertex parts which correspond to the 
coupling $\lambda {}_{0}$ of the duality model.  
The remaining part of the diagram consists of 
the low energy coherent part $\tilde{G}_{{\rm L}}$ or $\delta G$. \\
\indent We illustrate the decomposition of the 
skeleton diagrams of $\Phi [G]$ up to the 4th order in $U$  
following the above mentioned procedure in Fig.[6],
where the thick solid line and thin solid one 
represent $\tilde{G}_{{\rm H}}$, the high energy incoherent part, and
$\tilde{G}_{{\rm L}}$, the low energy coherent part,respectively,
and the dotted line represents the bare Coulomb potential $U$,
and the wavy line represents the spin fluctuation propagator $\chi_{0}$
which represents the longitudinal part $\chi_{0}^{z}$ or 
the transverse part $\chi_{0}^{+-}=\chi_{0}^{x}+\chi_{0}^{y}$.
In Fig.[7] we show the spin fluctuation propagator $\chi_{0}^{z}$
 and $\chi_{0}^{+-}$
in terms of $\tilde{G}_{{\rm H}}$ up to the second order in $U$.
We take up to the first correction of $\tilde{G}_{{\rm L}}$ by 
$\chi_{0}$. This is consistent with retaining the terms up to the
4th order in $U$ for $\Phi [G]$.
The coefficients are attached to the diagrams in Fig.[6]
because they are important when we compare with
those corresponding to the expansion in the duality model.
It is noted that the bubble of high energy component, in the second diagram of 
Fig.[6-a], includes both the charge susceptibility and the longitudinal 
spin susceptibility $\chi_{0}^{z}$ in this level of approximation.  However, 
since the charge fluctuations are irrelevant here, the coefficient of 
$\chi_{0}^{z}$ should be assigned so as to maintain the isotropy in the spin 
space by comparing with the diagram containing the transverse susceptibility 
$\chi_{0}^{+-}$, the first diagram of Fig.[6-a] for instance. \\
\indent Here, as an example, we explain the diagrams derived from the
$\frac{\delta{}^{2} \Phi}{\delta G_{1} \delta G_{2}}
|{}_{G=\tilde{G}{}_{{\rm H}}}$
{\small $\delta^{\prime} G_{1}\delta^{\prime} G_{2}$ }
of the diagram g in Fig [6].
The diagrams (g-1) and (g-2) are included in  
the diagram A in the expansion of
the duality model with the correct coefficient $1/2$.
Indeed, the diagram (g-1) can be regarded as a correction to the first term
of Fig.[6-a] which contains the first term of $\chi_{0}$ 
given in Fig.[7].  
The diagram (g-2) contains the two equivalent 
diagrams which give corrections 
of two vertices of the first term of Fig.[6-a] because 
the diagram A in the duality model has the spin-fermion coupling 
$\lambda {}_{0}$ at both edges of the spin fluctuation $\chi_{0}$.
The diagrams (g-4) and (g-5)  contain the first correction to those of 
(g-1) and (g-2) by the spin fluctuations included in
$\delta G$. The diagram (g-5) has two equivalent diagrams 
due to the same reason as for the diagram (g-2). \\
\indent Thus  the diagrams, derived from the diagrams of Fig.[1-a
 $\sim$ h]
for $\Phi[G]$ by the procedure 
$\frac{\delta {}^{2} \Phi}{\delta G_{1} 
\delta G_{2}}|{}_{G=\tilde{G}{}_{{\rm H}}}${\small $
\delta^{\prime} G_{1}\delta^{\prime} G_{2}$},
 can be rearranged into the
diagrams A,B,C and D of Fig.[2] in the duality model  with the coefficient
$1/2$, $1$, $1/4$ and  $1/4$, respectively.
Half of the term corresponding to the diagram B cancels with the diagrams 
of Fig.[4] and Fig.[5]. Indeed, when we sum up the diagrams in 
Fig.[4] $\sim$ [5], corresponding to the terms (2.12) and (2.13)
without the second term, up to the 4th order in the coupling 
$\lambda_{0}$, we are left with the diagram the same as B 
of the duality model with a coefficient $-1/2$.
So after all we can reproduce
the expansion of the thermodynamic potential in the duality model, shown
 in Fig.[2] up to the 4th order perturbation in the spin-fermion 
 coupling.
 This kind of comparison can be extended to higher order perturbation
 without difficulty in principle.
 It is left for future investigations to show that the above correspondence
 between the duality model and the strongly correlated original Hamiltonian
 hold to all orders of skeleton expansion of $\Phi [G]$
 in the intermediate stage of the renormalization.\\
 \indent In conclusion of this section, we have argued 
 on the basis of the renormalized 
perturbation approach that the thermodynamic potential of the Hubbard model 
with strong correlation can be reduced to that of the itinerant-localized 
duality model.  Although we have adopted the Hubbard model for conciseness 
of presentations, the argument can be extended without difficulty to other 
models such as the periodic Anderson model. 
\section{Spin-fluctuation propagator of  
 ``localized spin'' exhibiting superexchange-type correlation}
In this section we discuss a structure of the spin-fluctuation 
propagator $\chi_{0}(q,\omega)$ on the basis of an explicit calculation 
due to the 2nd order pertubation solution of the two-dimensional (2d) 
Hubbard model.  
The Hubbard Hamiltonian is written as
\begin{eqnarray}
 {\it H} &=& \sum_{k\sigma}\epsilon_{k}c^{\dagger}_{k\sigma}c_{k\sigma}
 + \frac{U}{N}
\sum_{k,k^{\prime}}\sum_{q(\neq 0)}c^{\dagger}_{k-q\uparrow}c^{\dagger}
_{k^{\prime}+q\downarrow}c_{k^{\prime}\downarrow}c_{k\uparrow},
\end{eqnarray}
where the conventional notations are used and  
$N$ is the number of site.
The explicit form of the one-particle energy $\epsilon_{k}$ 
is given by means of
 the hopping term $-t/2$ 
 between the nearest-neighbor sites as
\begin{eqnarray}
\epsilon_{k}&=&-t({\rm cos}{\it k_{x}a}+{\rm cos}{\it k_{y}a}),
\end{eqnarray}
 where {\it a} is the lattice constant. It is noted that the full bandwidth 
 {\it W} is equal to 4{\it t}.\\
\indent The spin susceptibility $\chi_{0}(q,{\rm i}\omega_{m})$ arising  from
the high energy processes is decomposed formally as
\begin{eqnarray}
\chi_{0}{}^{-1}(q,{\rm i}\omega_{m}) 
&=& \chi^{-1}({\rm i}\omega_{m}) -J(q,{\rm i}\omega_{m}),
\end{eqnarray}
where  $\chi({\rm i}\omega_{m})$ represents the one-site effect and
$J(q,{\rm i}\omega_{m})$  intersite effect among the
``localized-spin''  degrees of freedom.
If $\chi_{0}(q,{\rm i}\omega_{m})$ has a peak at the wavevector
$Q=(\pi /a, \pi/a)$, it means that
the nearest neighbor interaction is dominant. 
 By investigating a $U$-dependence of the peak height of $\chi_{0}(q,0)$, 
we can obtain a physical picture for the interaction between the 
``localized spins".  If $J(Q)$ scales as $1/U$ in the limit $U/W\gg 1$, 
we can conclude that it can be regarded as the superexchange interaction. \\
\indent In deriving the duality model in the previous section 
 we have set an important
assumption that the one particle spectral weight has
the triple peak structure. Therefore we have to calculate 
$\chi_{0}(q,{\rm i}\omega_{m})$, due to the high energy incoherent part,
 by means of the Green
functions which give the triple peak structure to the one particle spectrum.
It is well known that the 2nd order perturbation with respect to
$U$ reproduces the triple peaks structure. So we use 
 the 2nd order perturbed 
Green function $G^{(2)}(p,{\rm i}\epsilon_{n})$,
which includes the 2nd order perturbed 
 selfenergy $\Sigma^{(2)}(p,{\rm i}\epsilon_{n})$
 as an approximate Green function leading to  the duality model.
We further approximate $\chi_{0}$ as the first term shown in Fig.7.  
Namely, we take only the bubble diagram composed of 
$G^{(2)}_{\rm H}(p,{\rm i}\epsilon_{n})$, the incoherent part of 
$G^{(2)}(p,{\rm i}\epsilon_{n})$.  
The resultant structure of $\chi_{0}$ will be 
compared with the approximate form of the full susceptibility $\chi_{s}$ 
calculated in the same way but with the use of $G^{(2)}$ which contains the 
low energy contributions as well. \\
 \indent In order to take out precisely the high energy 
 part of the Green function,
 we perform the calculation on the real frequency axis.
The retarded selfenergy as a function
 of the coordinate $R$ and the time $t$ in
 the 2nd order perturbation with respect to  $U$ is given as~\cite{rf:Zlatic}
 \begin{eqnarray}
 \Sigma^{(2)}(R,t) & = & \frac{1}{{\rm i}}U^{2}[a^{2}(R,t)b(R,t) +
  a(R,t)b^{2}(R,t)],
 \end{eqnarray}
 where the function $a(R,t)$ and $b(R,t)$ are defined as
 \begin{subeqnarray}
  a(R,t)=\frac{1}{N}\sum_{p}{\rm e}^{{\rm i}(p\cdot R-\xi_{p}t)}f(\xi_{p}),\\
  b(R,t)=\frac{1}{N}\sum_{p}{\rm e}^{{\rm i}(p\cdot R-\xi_{p}t)}
  [1-f(\xi_{p})],
 \end{subeqnarray}
 where $\xi_{p} \equiv \epsilon_{p} -\mu$, $\mu$ being the
 chemical potential, and $f(\xi_{p})$ is the Fermi function.
 The function $a(R,t)$ and $b(R,t)$ can be calculated quickly by 
 the fast Fourier transformation (FFT) algorithm. 
 The selfenergy in the energy-momentum
 representation is given by
 \begin{equation}
 \Sigma (p,\omega) = \int_{0}^{\infty}{\rm d}t {\rm e}^{{\rm i}\omega t}
 \sum_{R}e^{-{\rm i}p\cdot R}\Sigma^{(2)}(R,t),
 \end{equation}
 where both the time integration and the momentum summation 
  are calculated by FFT.
 The calculations are performed with 128$\times$128 sites. We introduce
 the upper cut-off $\Lambda$ for the time integral
 and perform FFT in the interval $[0,\Lambda]$ 
 with  equally time spaced $N_{e}=1024$
 points. The cut-off $\Lambda$ restricts the energy interval to the region
 $[-D,D]$, where $D=\pi N_{e}/\Lambda$, and we set $\Lambda$ 
 so as to $D=5W=20t$ \\
 \indent The chemical potential $\mu$ is determined through the relation
 \begin{eqnarray}
  n &=& \frac{1}{N}\sum_{p}\int {\rm d}\epsilon\rho^{(2)}
(p,\epsilon)f(\epsilon),
 \end{eqnarray}
 where {\it n} is the electron number density per site and 
 $\rho^{(2)}(p,\epsilon)$ is
 the spectral weight defined as
 \begin{eqnarray}
  \rho^{(2)}(p,\epsilon) &\equiv& 
 -\frac{1}{\pi}{\rm Im} G^{(2)}(p,\epsilon) \nonumber \\
      &=& -\frac{1}{\pi}\frac{{\rm Im}\Sigma^{(2)} (p,\epsilon)}
 {[\epsilon-\mu-{\rm Re}\Sigma^{(2)} (p,\epsilon)]^{2}+
[{\rm Im}\Sigma^{(2)} (p,\epsilon)]^{2}}.
 \end{eqnarray}
 \indent In Fig[8a] and [8b] we show the calculated results of 
 the one-particle spectral weight of the systems 
 with $n=1.0$ and $n=0.7$, respectively,
 for various values of {\it U/W}.
 When the Coulomb interaction {\it U} is 
 comparable with the bandwidth {\it W},
 the one particle spectrum gets the lower and upper Hubbard bands 
 and becomes the triple peak structure, and exhibits the narrowing of 
 the coherent peak with increasing {\it U}.\\
 \indent We first calculate $\chi_{s}(q,\omega)$, since its calculation
 is simpler than that of $\chi_{0}(q,\omega)$ where the low energy 
 processes should be excluded.
 The approximate form of $\chi_{s}(q,\omega)$ 
 given by bubble diagram is calculated
 by  means of $G^{(2)}(p,\epsilon)$ as
 \begin{eqnarray}
 \chi_{s}(q,\omega) &=& 
 i\sum_{k}\int_{0}^{\infty}{\rm d}t{\rm e}^{{\rm i}\omega(t+\delta)}
[X^{\ast}{}_{k+q}(t)Y_{k}(t)-Y^{\ast}{}_{k+q}(t)X_{k}(t)].
 \end{eqnarray}
 where  $\delta$ is a positive infinitesimal and the function
 $X_{k}(t)$ and $Y_{k}(t)$ are defined as
 \begin{eqnarray}
 X_{k}(t) &\equiv& \int {\rm d}\epsilon {\rm e}^{{\rm i}t\epsilon}\rho^{(2)}
(k,\epsilon),\\
 Y_{k}(t) &\equiv& \int {\rm d}\epsilon {\rm e}^{{\rm i}t\epsilon}\rho^{(2)}
(k,\epsilon)
f(\epsilon).
 \end{eqnarray}
 The expression (3.9) is the convolution form of $X_{k}(t)$ and $Y_{k}(t)$
 in the  momentum
 space so that it can be calculated quickly by FFT.
 The calculation is performed at low enough temperature 
 {\it T}=0.01{\it t}=0.0025{\it W}.\\
 \indent The momentum dependence of the approximate full 
 susceptibility $\chi_{s}(p,\omega)$ for
 $\omega=0$ is shown
 in Fig[9a], [9b] and [9c] for various fillings.
 The incommensurate peaks can be seen for 
all value of {\it U/W} in the case of filling $n$=0.7 (Fig[9c]), 
 while they lose their height with increasing 
 the ratio of {\it U/W}. In the case $n$=0.8, 
 the incommensurate peaks disappear even for {\it U/W}=1.0.
 In these figures the incommensurate peaks appear
 in rather large doping for large {\it U/W}. 
 This is the effect of the approximation that we take but
 there is a physical reason that the coherent 
 peak which gives the incommensurate
 peak in the susceptibility is narrowed and its weight shifts to the high energy incoherent part
 which is the localized component, so the incommensurate peaks
 tend to vanish in strong Coulomb potential $U$.
 We also see that the value of susceptibility 
 decreases with increasing $U$. We explain
 these below with paying attention to the role of incoherent lower and upper Hub bard bands. \\
 \indent Next we calculate the susceptibility $\chi_{0}(q,{\rm i}\omega_{m})$ 
which
 is determined only by $G_{{\rm H}}$'s representing the high energy
 processes. So, we need an explicit form of $G_{{\rm H}}$.  
 For constructing the Green function $G_{{\rm H}}$,
 we introduce
 the cut-off which separates the low energy coherent peak and the high energy Hubbard bands.
 Of course the one particle spectrum of the Green function must be the triple peak structure
 for getting such cut-off. \\
 \indent Here we give $G_{{\rm H}}$ as
\begin{eqnarray}
 G_{{\rm H}}(p,{\rm i}\epsilon_{n}) &=& 
\int {\rm d}x \frac{\rho_{{\rm H}}(p,x)}
{x-{\rm i}\epsilon_{n}}, \nonumber \\
\end{eqnarray}
 where $\rho_{{\rm H}}(p,x)$ is
\begin{eqnarray}
 \rho_{{\rm H}}(p,x) &=& \rho (p,x)g(x,c_{1},c_{2}),
\end{eqnarray}
 where $\rho (p,x)$ is usual one-particle spectral function,
 $-\frac{1}{\pi}{\rm Im}G(p,x)$, and the function $g(x,c_{1},c_{2})$ is
 the cut-off function which cuts the  weight of the low energy coherent part
 that locating in the energy region $[c_{1}, c_{2}]$ as
\begin{eqnarray}
 g(x,c_{1},c_{2}) &=& 1+\frac{1}{{\rm e}^{\Gamma (x-c_{1})}+1}
-\frac{1}{{\rm e}^{\Gamma (x-c_{2})}+1},
\end{eqnarray}
 where the $\Gamma$ is a cut-off parameter and we set $\Gamma=5t$.
 In order to decide the values $c_{1}$ and $c_{2}$, 
 we select the saddle points between the
 the Hubbard peak and the coherent peak. In Fig[10] we show the
 $\rho_{{\rm H}}(x)=\sum_{p}\rho_{{\rm H}}(p,x)$ for the half filled
 (a) and the $n=0.7$ (b) case
 together with $\rho (x)=\sum_{p}\rho (p,x)$. \\
 \indent We give $\chi_{0}(q,{\rm i}\omega_{m})$ 
 the susceptibility of the incoherent part as 
 \begin{eqnarray}
 \chi_{0}(q,{\rm i}\omega_{m}) &=& -T\sum_{\epsilon_{n}}\sum_{k}
 G_{{\rm H}}(k+q,{\rm i}\epsilon_{n}+{\rm i}\omega_{m})
G_{{\rm H}}(k,{\rm i}\epsilon_{n}) \\
 &=& \sum_{k}\int {\rm d}x\int {\rm d}x^{\prime} \frac{\rho_{{\rm H}}(k+q,x)
 \rho_{{\rm H}}(k,x^{\prime})}{{\rm i}\omega_{m}
+x^{\prime}-x}[f(x)-f(x^{\prime})].
 \end{eqnarray}
  Non-zero contributions to  $\chi_{0}(q,{\rm i}\omega_{m})$
 arise from the integration regions of $x$ and $x^{\prime}$ where $x$ belongs
 to the upper Hubbard band and $x^{\prime}$ to the lower one, or vice versa, 
 because the temperatures in question are low enough compared to the energy 
 cut-off separating the coherent and incoherent parts.
 Then we can roughly estimate the denominator ${\rm i}\omega_{m}+x^{\prime}-x$ 
 in (3.16) for low temperature as $1/U$ 
 and the difference of two Fermi distributions as $\pm 1$. 
 This seems to give $t^{2}/U$ expected for the  
 superexchange interaction. Unfortunately we cannot get the explicit 
 $1/U$ scaling for the peak height of $\chi_{0}(q,0)$ in our approximation,
 although $\chi_{0}(Q,0)$ is a decreasing function of $U$ in consistent
 with a general trend expected from the expression (3.16).\\
 \indent In Fig.[11a] we show the susceptibility of the
 incoherent parts $\chi_{0}(q,0)$, (3.16),
 for $n$=0.7 and $U/W$=1.0. It is noted that the peak of
 of $\chi_{0}(q,0)$ is located at commensurate position $Q=(\pi/a,\pi/a)$.
 This means that the spin degrees of freedom from the incoherent parts 
 interact most strongly between nearest neighbor sites.
 The same point of view about the superexchange interaction was
 suggested by Ohkawa {\it et al.}~\cite{rf:Ohkawa} 
 on the basis of  the {\it 1/d} expansion and rather rough estimate of 
 the incoherent parts.
 In our approximation scheme it is apparent that the superexchange-type 
 interaction is derived through 
 the electron-hole pair excitations between the lower and upper Hubbard bands.
 The spin susceptibility $\chi_{0}(q,\omega)$ appearing in the duality model 
 contains this information.\\
 \indent In Fig.[11b], we show the $q$-dependence of the full susceptibility 
$\chi_{\rm s}(q,0)$ calculated by (3.9).  It is noted that the peak 
of $\chi_{\rm s}(q,0)$ is located at the incommensurate position.  
This is due to the effect of coherent motion of quasiparticle component
which reflects the shape of the Fermi surface.  It is a general characteristic 
of the duality description that the susceptibility $\chi_{0}(q)$, composed of 
the incoherent component, has commensurate correlation while the full 
susceptibility $\chi_{\rm s}(q)$, taking into account the coupling with the 
fermions, has the incommensurate correlation in general.
 Indeed the full dynamical spin fluctuation propagator 
 in the duality model is formally
 written as 
 \begin{equation}
 \chi_{s}{}^{-1}(q,{\rm i}\omega_{m})=\chi_{0}{}^{-1}(q,{\rm i}\omega_{m})-
 \lambda_{0}{}^{2}\Pi (q,{\rm i}\omega_{m}),
 \end{equation}
 where the polarization $\Pi (q,{\rm i}\omega_{m})$ of the fermion component is
 given by
 \begin{equation}
 \Pi (q,{\rm i}\omega_{m})=-T\sum_{\epsilon_{n} p} \tilde{G}_{{\rm L}}(q+p,
 {\rm i}\omega_{m}+{\rm i}\epsilon_{n})
\tilde{G}_{{\rm L}}(p,{\rm i}\epsilon_{n}).
 \end{equation}
 Then $\chi_{s}(q,{\rm i}\omega_{m})$ exhibits peaks at incomensurate positions
 in general, reflecting the $q$-dependence of the polarization 
 $\Pi (q,{\rm i}\omega_{m})$.\\
 \indent It is noted that the incommensurate 
 correlation of $\chi_{\rm s}(q,\omega)$ 
 disappears as the frequency $\omega$ exceed the cut-off, separating 
 the coherent and incoherent components, leaving only the commensurate 
 correlation due to $\chi_{0}(q,\omega)$.  Such a behavior 
 has been recently discovered in by neutron scattering 
 in Cr,~\cite{rf:Fukuda}
 and will be discussed 
 in detail elsewhere.
 \section{Conclusion and discussion}
 In this paper we  have examined the duality picture 
 diagramatically and
 the origin of the superexchange interaction as electron hole pairs 
 excitation between lower and upper Hubbard bands. The central assumption 
 about the justification of the duality model
 is that in the intermediate stage of the renormalization, 
 where the duality model
 can be applied, the one particle spectrum already have 
 the triple peak structure  and the
 spin fluctuation process from the incoherent parts is relatively larger than other processes,
 for example than the charge fluctuation. Due to the explicit separation between the
 incoherent/localized and coherent/itinerant parts we obtain 
 naturally the ``localized-spin'' degree
 of freedoms with neglecting the other fluctuation freedoms. 
 Here a question arises whether we can actually separate the coherent 
 and incoherent parts so explicitly. In the impurity model
 like Anderson model and in the $d=\infty$ lattice model, $d$ 
 being the space dimension,
  which is equivalent to the impurity model
 under some conditions
 the coherent peak of the one particle spectrum is narrowed exponentially
 and the lower and upper Hubbard bands is developed
 explicitly with increasing the on site Coulomb $U$, 
 we can safely set the above assumption.
 In the $d=2$ and $d=3$ lattice systems case like Hubbard model, 
 the separation of the coherent
 and incoherent parts is not so clear as in the impurity cases.
 But we believe the duality model can apply to 
 these cases as an effective action
 when the electron correlation is so strong.\\
 \indent In $\S2$ we treat the fixed point just like as Fermi liquid but
 the same procedure is
 applicable for the non-Fermi liquid fixed point
 if the assumptions we mentioned are
 satisfied and there no singularity appears from the diagrams
 which we neglected.
 Actually we do not know the correct analytic form of the
 partially renormalized Green function
 $\tilde{G}(p,{\rm i}\epsilon_{n})$ 
and the susceptibility of the localized spin
 $\chi_{0}(q,{\rm i}\omega_{m})$, so we must
 input appropriate form of $\tilde{G}(p,{\rm i}\epsilon_{n})$ 
 and $\chi_{0}(q,\omega)$
 which suitably describe the physical situation we deal.
 For example in the reference 5) $\sim$ 7) the weak antiferromagnatism
 of heavy fermion and 
 the normal state properties of cupurate superconductor are successfully
 explained
 with the suitable
 form of $\tilde{G}(p,{\rm i}\epsilon_{n})$ which
 include the nesting property.
 But by considering the structures of the incoherent part of fermions which
 are hidden, for example, in the Fermi Liquid theory we get the another
 properties and explanation of the strongly correlated fermion systems.
\section*{Acknowledgements}
One of the authors (K.M.) has much benefitted from enlightening discussions
with Y. Kuramoto at early stage of this work.  This work is supported by the
Grant-in-Aid for Scientific Research (07640477), and the Grand-in-Aid for
Scientific Research on Priority Areas
``Physics of Strongly Correlated Conductors" (06244104) and
``Anomalous Metallic State near the Mott Transition" (07237102)
from the Ministry of Education, Science, Sports and Culture.

\newpage
\begin{center}
{\bf Figure Captions}
\end{center}
\begin{itemize}
\item Fig. [1] \newline
\indent Diagrammatic expansion of the Luttinger-Ward functional $\Phi [G]$
 up to the 4th order in the on-site Coulomb interaction.
The solid line represents the fully renormalized 
 Green function $G$ and the dotted line the bare Coulomb interaction $U$. 
 Summation of the momentum, frequency and spin index
 are abbreviated.\\

\item Fig. [2] \newline
\indent Diagrammatic expansion of the thermodynamic potential of the 
 duality model. The solid line represents the partially renormalized 
 Green function $\bar{G}$, and the wavy line the spin fluctuation
 propageter $\chi_{0}$, and the dot the spin-fermion coupling
 $\lambda_{0}$.\\

\item Fig. [3] \newline
\indent Diagrammatic expression of $\delta G_{\sigma}(p,{\rm i}\epsilon_{n})$.
 The wavy line represents  $\chi_{0}$ and the dot-dashed line represents  
 $\tilde{G}_{{\rm L}}$, the dot is the spin-fermion vertex part which
 corresponds  $\lambda_{0}$ of the duality model. \\

\item Fig. [4] \newline
\indent Diagrammatic expression of $(\delta G/\tilde{G})$ 
and $(\delta G/\tilde{G})^{2}$ \\

\item Fig. [5] \newline
\indent Diagramatic expression of $(\delta \Sigma \tilde{G}_{{\rm L}} + 
 \delta \Sigma \delta G)$ in (2.13).\\

\item Fig. [6] \newline
\indent Decomposition of the skeleton diagrams 
 for the thermodynamic potential $\Phi[G]$ into the diagrams 
 with high and low energy component.
 The thin line represents $\tilde{G}_{{\rm L}}$, the low energy part
 of the partially renormalized Green function and the thick line 
 $\tilde{G}_{{\rm H}}$, the high energy one, and 
 the dotted line  the bare Coulomb interaction {\it U}.
 The symbol $\in$ means that the diagrams are included in  
 the corresponding diagram $({\rm A} \sim {\rm D})$ of the duality model
 shown at the left side.\\

\item Fig. [7] \newline
\indent Diagrammatic expression of $\chi_{0}^{z}(q,\omega)$
 and $\chi_{0}^{+-}(q,\omega)$. The solid line 
 represents $\tilde{G}_{{\rm H}}$, the high energy part of the partially 
 renormalized Green function, and the dotted line the bare Coulomb
 interaction  {\it U}.  \\

\item Fig. [8] \newline
\indent One-particle spectral weight $\sum_{p}\rho^{(2)}(p,\epsilon)$,
 given by (3.8), 
 of the 2d Hubbard model 
 for various values of 
 $U/W$ with the filling (a) $n$=1.0 and (b) $n$=0.7. \\

\item Fig. [9] \newline
\indent Wavevector dependence 
 of the approximate full susceptibility $\chi_{s}(q,0)$ 
for various values of {\it U/W}. The filling are (a) {\it n}=1.0,
 (b) {\it n}=0.8, and (c) {\it n}=0.7, respectively.\\

\item Fig. [10] \newline
\indent One particle spectral weight $\rho_{\rm H}(\omega)$, the incoherent
 part which is indicated by the
 solid line, and $\rho^{(2)}(\omega)=\sum_{p}\rho^{(2)}(p,\omega)$, given
 by (3.8), which is indicated by the dotted line,
 for the case  $U/W$=1.0. The fillings are (a) {\it n}=1.0 
 and (b) {\it n}=0.7\\
 
\item Fig. [11] \newline
\indent The wavevector dependence of (a)
 the susceptibility $\chi_{0}(q,0)$ of the incoherent part
 and (b) the full susceptibility $\chi_{s}(q,0)$
 for the case  $U/W$=1.0 and  $n$=0.7\\
 
\end{itemize}
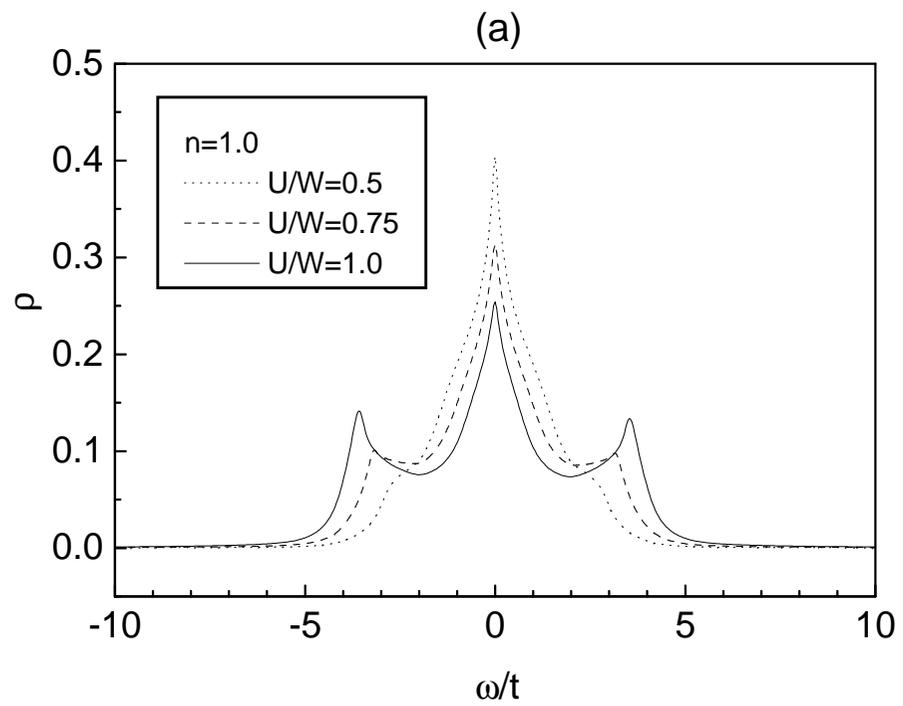
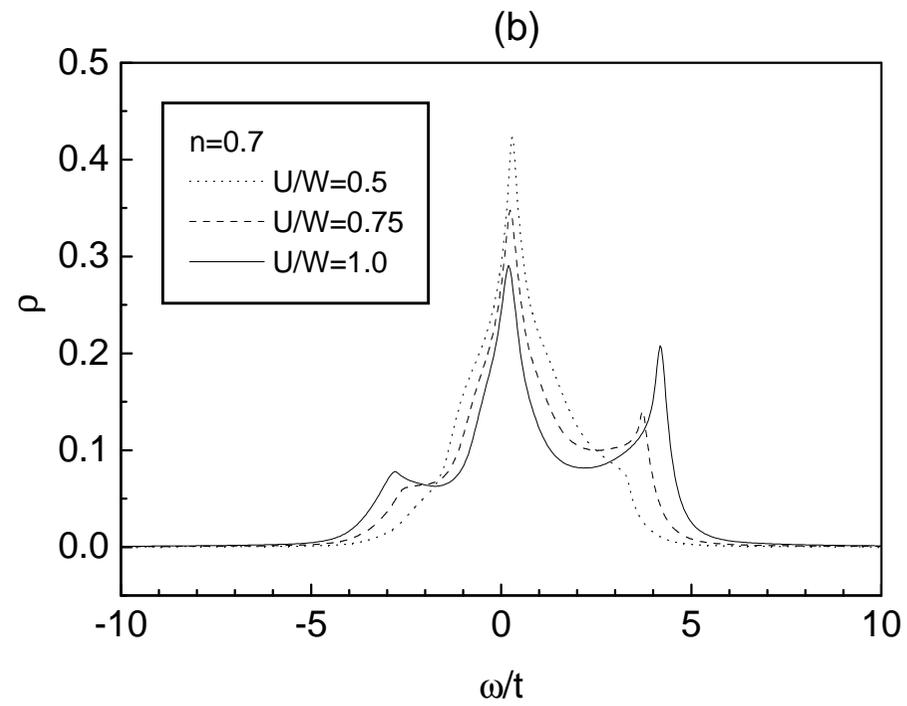
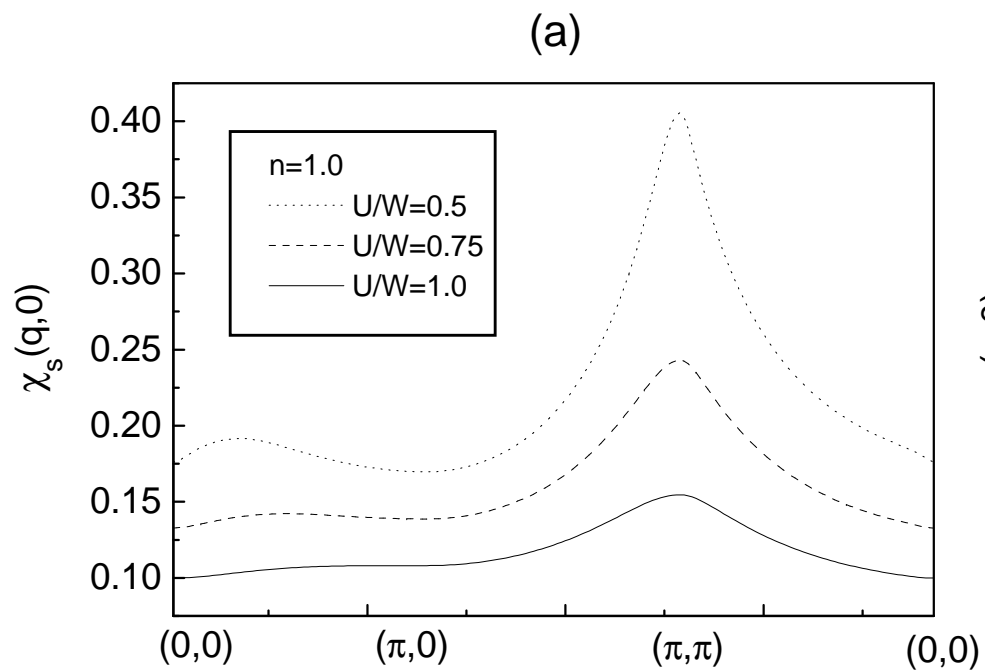
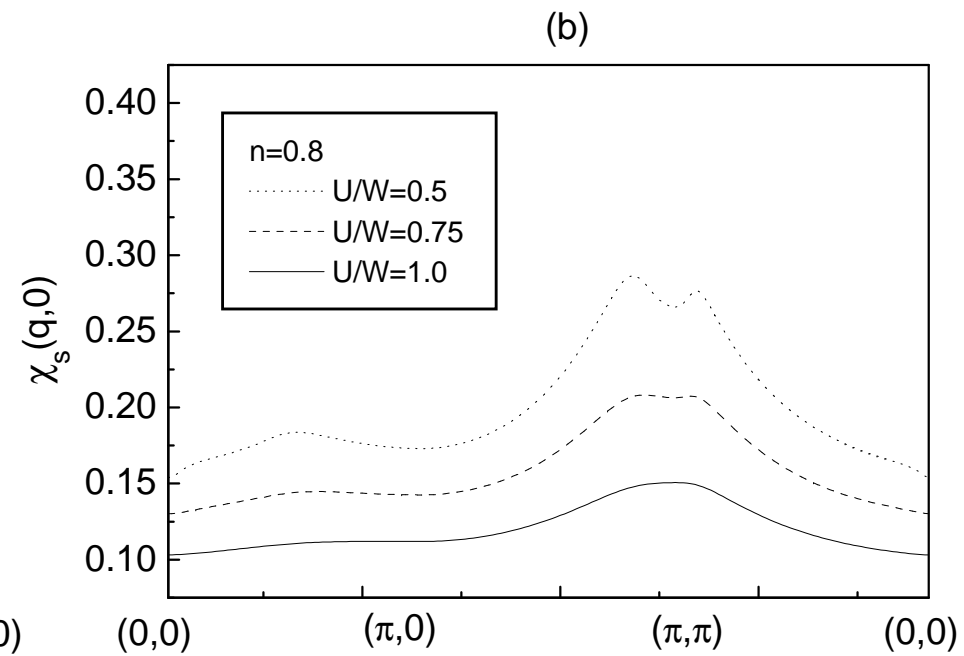
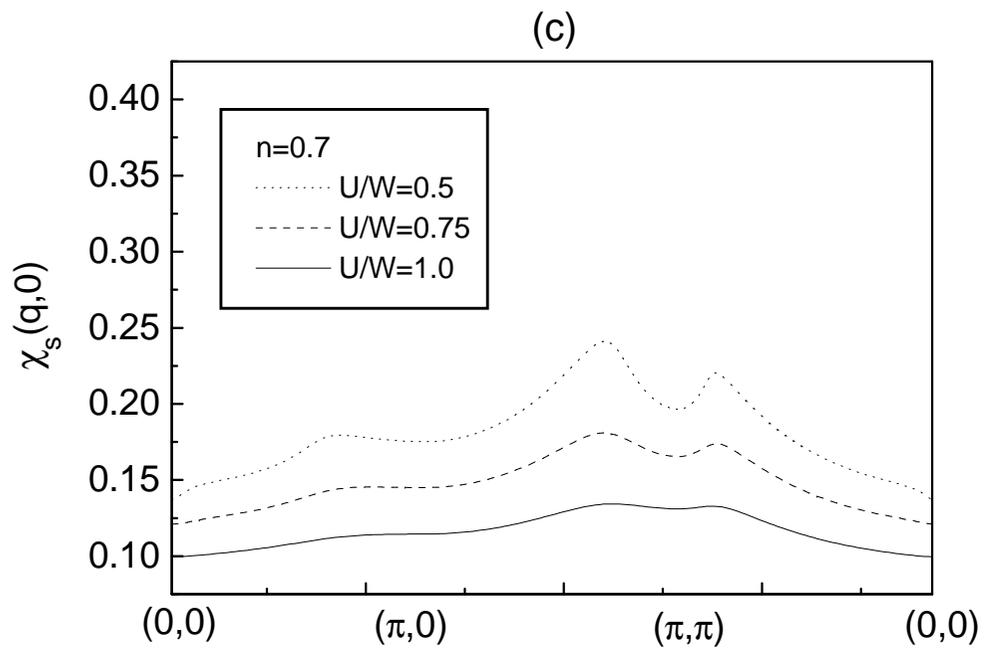
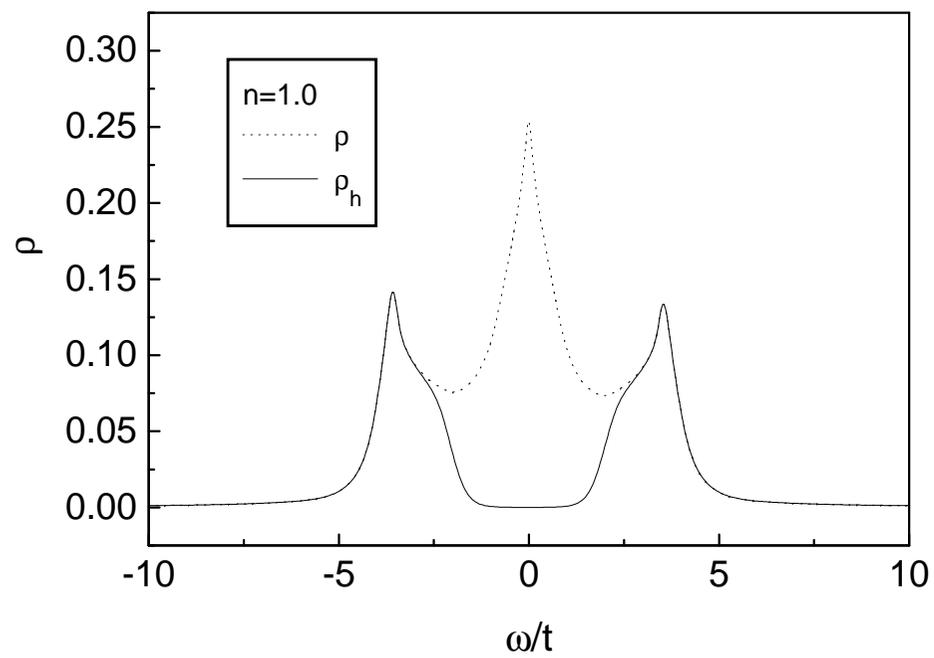
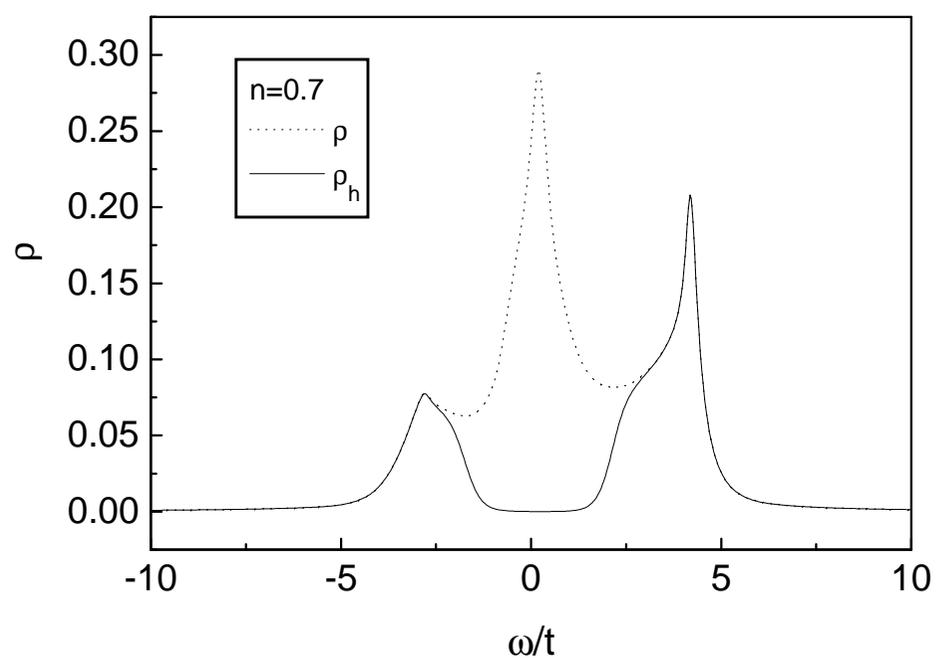
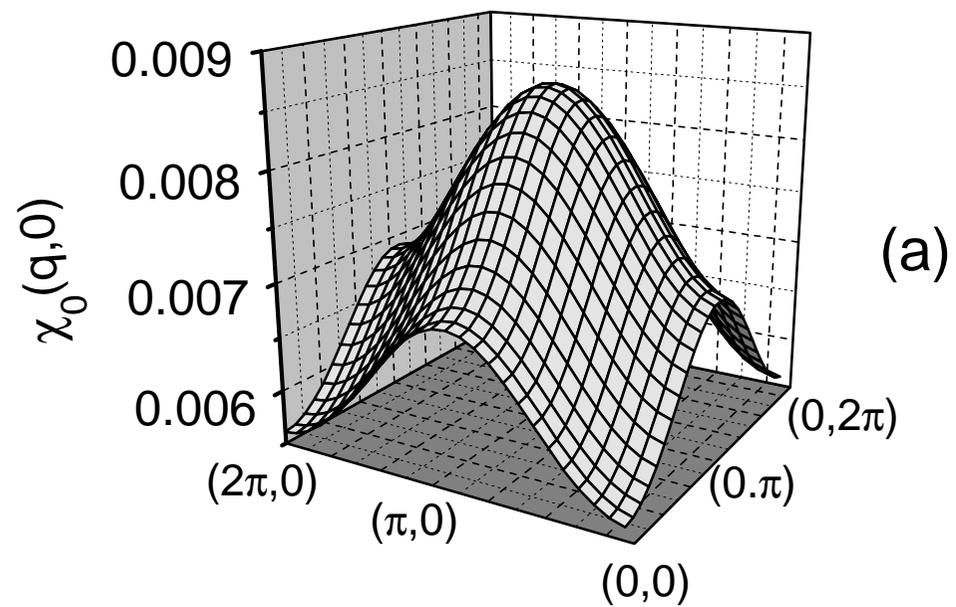
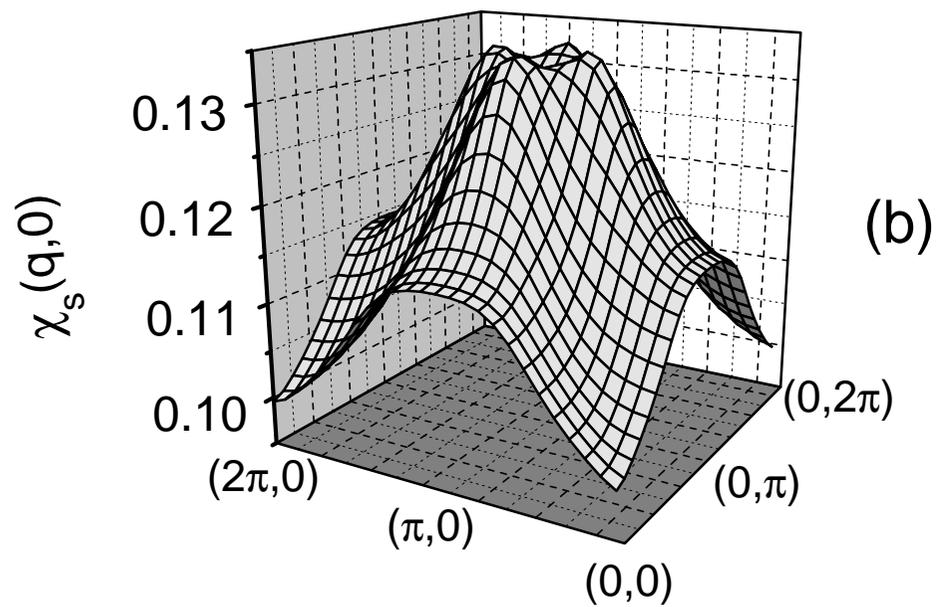
\end{document}